\documentstyle[aps,epsfig]{revtex}
\begin{document}

\title{ Neutron and proton drip lines using the modified Bethe-Weizs\"acker mass formula }

\author{D.N. Basu\thanks{E-mail:dnb@veccal.ernet.in}}
\address{Variable  Energy  Cyclotron  Centre,  1/AF Bidhan Nagar,
Kolkata 700 064, India}
\date{\today }
\maketitle
\begin{abstract}

      Proton and neutron separation energies have been calculated using the extended Bethe-Weizs\"acker mass formula. This modified Bethe-Weizs\"acker mass formula describes minutely the positions of all the old and the new magic numbers. It accounts for the disappearance of some traditional magic numbers for neutrons and provides extra stability for some new neutron numbers. The neutron and proton drip lines have been predicted using this extended Bethe-Weizs\"acker mass formula. The implications of the proton drip line on the astrophysical rp-process and of the neutron drip line on the astrophysical r-process have been discussed. 

 

\end{abstract}



\section{Introduction}
\label{sectoin1}

      Recent improvements of radioactive ion beam (RIB) technology allow measurements of the masses, half-lives, radii and other properties of unstable nuclei. Nuclear reactions using RI beams have also been studied extensively. Based on these results, some interesting characteristics of unstable nuclei, such as a halo and a skin \cite{r1}, have been revealed. The disappearance of some traditional magic numbers and extra stability for some neutron numbers \cite{r2} have also been observed. The  Bethe-Weizs\"acker (BW) mass formula \cite{r3} designed to reproduce the gross features of nuclear binding energies for medium and heavy nuclei fails for light nuclei, specially away from the line of stability. The unusual stability of nuclei with preferred nucleon numbers commonly referred to as magic numbers and have been explained to be due to nuclear shell structure can be clearly delineated by the Bethe-Weizs\"acker mass formula. But the newly observed features like the disappearance of some traditional magic numbers and extra stability for some neutron numbers can not be identified.

      In the present work the one neutron and one proton separation energies have been calculated using the newly extented Bethe-Weizs\"acker (BW) mass formula \cite{r4} which delineates the positions of all old and the new magic numbers. This modified BW mass formula also explains the shapes of the binding energy versus neutron number curves of all the elements from $Li$ to $Bi$. For heavier nuclei this modified BW mass formula approaches the old BW \cite{r3} mass formula. The nucleus from which removal of a single neutron (and any more) makes the one proton separation energy negative defines a proton drip line nucleus and the nucleus to which addition of a single neutron (and any more) makes the one neuton separation energy negative defines a neutron drip line nucleus. Proton and neutron drip line nuclei, thus obtained, over the entire (N,Z) region, where N and Z are the neutron and atomic numbers, respectively,  of nuclei, define the predicted neutron and proton drip lines. Since the new mass formula has been capable of identifying all the new magicities or its loss, calculations according to this mass formula are expected to provide better limits to the neutron and proton drip lines.       

      Investigations near N=Z reveals the role of the neutron-proton interaction and its consequences for p-n pairing. Steep decrease in the isoscalar p-n pairing energy has been suggested with increasing $\mid N-Z \mid$ \cite{r5}. Moreover, the light neutron rich nuclei away from the region of $\beta$-stability have been found to be more bound than the old BW mass formula predictions indicating the existence of a more complicated dependence of the nuclear binding energy on the neutron-proton asymmetry. The symmetry energy coefficient which has been found to decrease with decreasing density for the asymmetric nuclear matter \cite{r6} suggests that for light nuclei with relatively large low density region near the surface, the effective symmetry energy coefficient should be smaller compared to heavier nuclei. These observations warrant the need for modifications of the old BW mass formula encompassing the domain of light nuclei that can help to identify the new magicity or, its loss and, give a better limit to the neutron drip line.   

\section{The modified Bethe-Weizs\"acker mass formula }
\label{section2}

      The binding energy of any nucleus of mass number A and atomic number Z obtained from a phenomenological search can be given by a more genaralized BW formula \cite{r4} 

\begin{equation}
 B(A,Z) = a_vA-a_sA^{2/3}-a_cZ(Z-1)/A^{1/3}-a_{sym}(A-2Z)^2/[(1+e^{-A/k})A]+\delta_{new},
\label{seqn1}
\end{equation}
\noindent
where the asymmetry term and the pairing term have been modified to obtain great improvements to the fits of the binding enegy versus neutron number curves of the light nuclei. Fitted value of the asymmetry term modifying constant $k=17$, while the modified pairing term $\delta_{new}$ can be expresssed as  

\begin{equation}
 \delta_{new}=(1-e^{-A/c})\delta~~~~where~~~~c=30,
\label{seqn2}
\end{equation}
\noindent
and $\delta$ being the old pairing term is given by

\begin{eqnarray}
  \delta=&&a_pA^{-1/2}~for~even~N-even~Z,\nonumber\\
         =&&-a_pA^{-1/2}~for~ odd~N-odd~Z,\nonumber\\
         =&&0~for~odd~A,\nonumber\\
\label{seqn3}
\end{eqnarray} 
\noindent
while the other constants

\begin{equation}
 a_v=15.79~MeV,~a_s=18.34~MeV,~a_c=0.71~MeV,~a_{sym}=23.21~MeV~and~a_p=12~MeV,
\label{seqn4}
\end{equation}
\noindent
remaining the same as the corresponding values of the constants for the old BW formula. Such modifications, however, do not alter significantly the results for heavier nuclei. 

      This modified BW formula suggests additional stabilities for few light nuclei at neutron numbers N=16 (Z=7,8), N=14 (Z=7-10), Z=14 (N=13-19), N=6 (Z=3-8) and loss of magicity for nuclei with neutron numbers N=8 (Z=4) and N=20 (Z=12-17), most of which were suggested earlier from the systematics. The magicity at neutron number N=6 (Z=3-8) suggested by the new formula is also supported by the experimental r.m.s. radii values that show a quenching at N=6 (Z=3-8). The new formula supports already known $^{32}Ne$ \cite{r7}, $^{35}Na$ \cite{r8}, $^{38}Mg$ \cite{r9}, $^{41}Al$ \cite{r9}, as the last bound isotopes of neon, sodium, magnesium and aluminium, respectively. Wherever there is a signature of shell closure in experimental data, the BW as well as the new modified BW formula both show marked deviations delineating clearly the positions of old magic numbers at 2, 8, 20, 28, 50, 82 and 126. 

\section{The neutron and proton separation energies and the drip lines}
\label{section3}

      The one proton separation energy $S_p$, which is the energy required to remove one proton from a nucleus is given by

\begin{equation}
 S_p = B(A,Z) - B(A-1,Z-1),
\label{seqn5}
\end{equation} 
\noindent
while the one neutron separation energy $S_n$, which is the energy required to remove one neutron from a nucleus is   given by

\begin{equation}
 S_n = B(A,Z) - B(A-1,Z),
\label{seqn6}
\end{equation}
\noindent
where B(A,Z), B(A-1,Z) and B(A-1,Z-1) are the binding energies of nuclei having mass number and atomic number A,Z, A-1,Z and A-1,Z-1 respectively.
 
      The one proton and one neutron separation energies have been calculated according to Eq.(5) and Eq.(6), respectively, using the new modified BW mass formula given by Eq.(1) with the $\delta_{new}$ quantity described by Eq.(2). The best fit values used for the new constants are $k=17$ and $c=30$ while the values of the older constants used for calculating the binding energies are given in Eq.(4). Then for a fixed value of the atomic number Z, starting from a very low value of neutron number N, which has been increased in steps of one till the one proton separation energy $S_p$ becomes  positive for the first time and the proton drip line nucleus is reached. Similarly, again for a fixed value of the atomic number Z, starting from a very high value of neutron number N, which has been decreased in steps of one till the one neutron separation energy $S_n$ becomes  positive for the first time and the neutron drip line nucleus is reached.   

\begin{figure}[h]
\eject\centerline{\epsfig{file=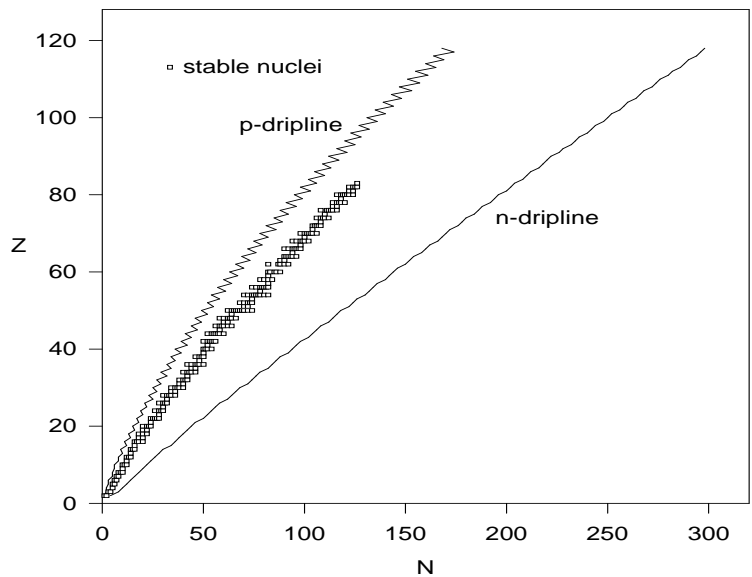,height=14cm,width=14cm}}
\caption
{The atomic number Z has been plotted against the neutron number N for the stable nuclei and the driplines calculated using the modified BW mass formula.}
\label{fig1}
\end{figure}

\begin{figure}[h]
\eject\centerline{\epsfig{file=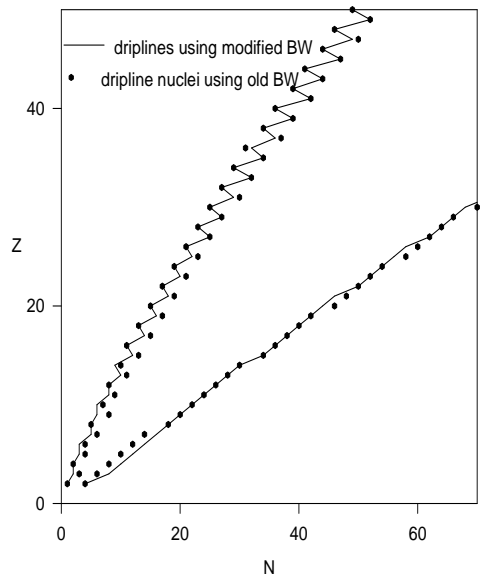,height=14cm,width=9cm}}
\caption
{The atomic number Z has been plotted against the neutron number N for the driplines calculated using the modified BW mass formula and the dripline nuclei using the old BW mass formula.}
\label{fig2}
\end{figure}

      In Fig.~\ref{fig1}, using the new modified BW mass formula, the atomic number Z has been plotted against the neutron number N for the proton and neutron dripline nuclei. The lines joining the proton dripline nuclei and the neutron dripline nuclei have been shown as the p-dripline and the n-dripline respectively in the figure. The nuclei stable against decay via $\beta^{\pm}$, $\alpha$ or heavy particle emissions or electron capture or spontaneous fission \cite{r10} have been shown with small hollow squares. There is no nucleus beyond the atomic number Z=83 which is stable against the decays via $\beta^{\pm}$, $\alpha$, heavy particle emissions and electron capture and spontaneous fission. In the Table-1 below, using the new modified BW mass formula, the atomic number Z and the neutron number N of the neutron and proton drip line nuclei have been tabulated along with the corresponding values of their one proton and one neutron separation energies. The formula wrongly predicts $^5Li$, $^6Be$, $^{15}F$ and $^{16}Ne$ as stable, and puts the neutron drip line at $^{17}B$, $^{20}C$ and $^{26}O$, while $^{19}B$, $^{22}C$ are in fact also stable, but $^{25,26}O$ are not. Except for these few exceptions where the Bethe-Weizs\"acker mass formula could not be as well tuned as predicting such fine details, the values given in Table-1 could be useful for planning experiments on mass measurements. In order to demonstrate the change of sign of the one proton and one neutron separation energies, the atomic number Z and the neutron number N of nuclei with one neutron less than the proton drip line nuclei and nuclei with one neutron beyond the neutron drip line nuclei have also been listed along with the corresponding values of their one proton and one neutron separation energies. In the seventh column of Table-1, the predicted values of the atomic and the neutron numbers (Z,N) for the last bound isotopes of neutron rich nuclei have been listed. In Fig.~\ref{fig2}, using the old BW mass formula which can be obtained by putting $k=0$ and $c=0$ in Eq.(1) and Eq.(2) respectively, the atomic number Z and the neutron number N of the neutron and proton drip line nuclei along with proton and neutron driplines using the new modified BW mass formula have been plotted. The modified BW mass formula alters significantly the results of the low mass region only. The nuclei beyond $Z=47$ the proton drip line using the old BW and modified BW are identical while for nuclei beyond $Z=30$ the neutron drip line using the old BW and modified BW are identical. However, the proton and the neutron separation energies still differ and gradually becomes identical for heavier nuclei.   

\begin{table}
\caption{Proton and neutron separation energies on and just beyond the proton and the neutron drip lines using the new modified BW mass formula.}
\begin{tabular}{cccccccccccc}
Proton& $S_p$&$S_n$ &One& $S_p$&$S_n$ &Neutron& $S_p$&$S_n$ &One& $S_p$&$S_n$   \\
&&&beyond&&&&&&beyond&& \\
&&&proton&&&&&&neutron&& \\ 
Drip    &         &           &drip &           &          &Drip     &          &          &drip&          &      \\ \hline
Z,N  &$ MeV$ & $MeV$&Z,N  &$ MeV$ & $MeV$ &Z,N  &$ MeV$ & $MeV$ &Z,N  &$ MeV$ & $MeV$  \\ \hline
  2,  1&  .211E+01&  .267E+02&  2,  0& -.819E+01&  .888E+02&  2,  4&  .238E+02&  .337E+01&  2,  5&  .278E+02& -.475E+00\\
  3,  2&  .337E+01&  .196E+02&  3,  1& -.348E+01&  .322E+02&  3,  8&  .270E+02&  .598E+00&  3,  9&  .293E+02& -.271E+01\\
  4,  2&  .112E+01&  .249E+02&  4,  1& -.415E+01&  .359E+02&  4, 10&  .283E+02&  .918E+00&  4, 11&  .304E+02& -.249E+01\\
  5,  3&  .754E+00&  .197E+02&  5,  2& -.325E+01&  .289E+02&  5, 12&  .255E+02&  .998E+00&  5, 13&  .273E+02& -.246E+01\\
  6,  3&  .187E+00&  .232E+02&  6,  2& -.331E+01&  .323E+02&  6, 14&  .278E+02&  .103E+01&  6, 15&  .295E+02& -.245E+01\\
  7,  5&  .199E+01&  .169E+02&  7,  4& -.887E+00&  .231E+02&  7, 16&  .251E+02&  .987E+00&  7, 17&  .266E+02& -.249E+01\\
  8,  5&  .199E+01&  .196E+02&  8,  4& -.698E+00&  .259E+02&  8, 18&  .278E+02&  .953E+00&  8, 19&  .292E+02& -.251E+01\\
  9,  6&  .107E+00&  .210E+02&  9,  5& -.231E+01&  .220E+02&  9, 20&  .250E+02&  .907E+00&  9, 21&  .263E+02& -.253E+01\\
 10,  6&  .649E+00&  .233E+02& 10,  5& -.165E+01&  .243E+02& 10, 22&  .278E+02&  .878E+00& 10, 23&  .290E+02& -.253E+01\\
 11,  8&  .549E+00&  .200E+02& 11,  7& -.148E+01&  .200E+02& 11, 24&  .249E+02&  .856E+00& 11, 25&  .261E+02& -.251E+01\\
 12,  8&  .139E+01&  .220E+02& 12,  7& -.574E+00&  .220E+02& 12, 26&  .277E+02&  .849E+00& 12, 27&  .288E+02& -.248E+01\\
 13, 10&  .725E+00&  .194E+02& 13,  9& -.104E+01&  .188E+02& 13, 28&  .249E+02&  .854E+00& 13, 29&  .259E+02& -.243E+01\\
 14,  9&  .803E-01&  .206E+02& 14,  8& -.165E+01&  .256E+02& 14, 30&  .276E+02&  .868E+00& 14, 31&  .285E+02& -.237E+01\\
 15, 12&  .754E+00&  .190E+02& 15, 11& -.810E+00&  .181E+02& 15, 34&  .265E+02&  .567E-01& 15, 35&  .274E+02& -.308E+01\\
 16, 11&  .477E+00&  .196E+02& 16, 10& -.107E+01&  .243E+02& 16, 36&  .291E+02&  .136E+00& 16, 37&  .299E+02& -.296E+01\\
 17, 14&  .696E+00&  .187E+02& 17, 13& -.714E+00&  .175E+02& 17, 38&  .261E+02&  .222E+00& 17, 39&  .269E+02& -.284E+01\\
 18, 13&  .706E+00&  .189E+02& 18, 12& -.691E+00&  .234E+02& 18, 40&  .286E+02&  .300E+00& 18, 41&  .294E+02& -.272E+01\\
 19, 16&  .582E+00&  .185E+02& 19, 15& -.702E+00&  .171E+02& 19, 42&  .258E+02&  .385E+00& 19, 43&  .265E+02& -.259E+01\\
 20, 15&  .820E+00&  .184E+02& 20, 14& -.456E+00&  .227E+02& 20, 44&  .282E+02&  .459E+00& 20, 45&  .289E+02& -.248E+01\\
 21, 18&  .430E+00&  .183E+02& 21, 17& -.750E+00&  .168E+02& 21, 46&  .253E+02&  .540E+00& 21, 47&  .260E+02& -.235E+01\\
 22, 17&  .852E+00&  .180E+02& 22, 16& -.322E+00&  .221E+02& 22, 50&  .289E+02&  .814E-01& 22, 51&  .295E+02& -.273E+01\\
 23, 20&  .250E+00&  .181E+02& 23, 19& -.840E+00&  .165E+02& 23, 52&  .261E+02&  .178E+00& 23, 53&  .267E+02& -.260E+01\\
 24, 19&  .821E+00&  .176E+02& 24, 18& -.264E+00&  .217E+02& 24, 54&  .284E+02&  .260E+00& 24, 55&  .289E+02& -.248E+01\\
 25, 22&  .510E-01&  .180E+02& 25, 21& -.962E+00&  .163E+02& 25, 56&  .256E+02&  .348E+00& 25, 57&  .261E+02& -.236E+01\\
 26, 21&  .743E+00&  .173E+02& 26, 20& -.266E+00&  .213E+02& 26, 58&  .278E+02&  .422E+00& 26, 59&  .283E+02& -.225E+01\\
 27, 25&  .773E+00&  .142E+02& 27, 24& -.164E+00&  .179E+02& 27, 62&  .261E+02&  .901E-01& 27, 63&  .266E+02& -.251E+01\\
 28, 23&  .627E+00&  .171E+02& 28, 22& -.316E+00&  .209E+02& 28, 64&  .282E+02&  .168E+00& 28, 65&  .287E+02& -.240E+01\\
 29, 27&  .489E+00&  .142E+02& 29, 26& -.391E+00&  .178E+02& 29, 66&  .255E+02&  .251E+00& 29, 67&  .260E+02& -.229E+01\\
 30, 25&  .481E+00&  .169E+02& 30, 24& -.403E+00&  .206E+02& 30, 68&  .276E+02&  .321E+00& 30, 69&  .280E+02& -.219E+01\\
 31, 29&  .202E+00&  .142E+02& 31, 28& -.628E+00&  .177E+02& 31, 72&  .259E+02&  .500E-01& 31, 73&  .263E+02& -.240E+01\\
 32, 27&  .310E+00&  .167E+02& 32, 26& -.521E+00&  .204E+02& 32, 74&  .278E+02&  .121E+00& 32, 75&  .282E+02& -.230E+01\\
 33, 32&  .676E+00&  .161E+02& 33, 31& -.884E-01&  .142E+02& 33, 76&  .253E+02&  .197E+00& 33, 77&  .257E+02& -.220E+01\\
 34, 29&  .119E+00&  .166E+02& 34, 28& -.665E+00&  .201E+02& 34, 78&  .273E+02&  .262E+00& 34, 79&  .276E+02& -.210E+01\\
 35, 34&  .343E+00&  .161E+02& 35, 33& -.381E+00&  .142E+02& 35, 82&  .255E+02&  .310E-01& 35, 83&  .259E+02& -.228E+01\\
 36, 32&  .635E+00&  .183E+02& 36, 31& -.881E-01&  .165E+02& 36, 84&  .274E+02&  .958E-01& 36, 85&  .278E+02& -.219E+01\\
 37, 36&  .131E-01&  .161E+02& 37, 35& -.674E+00&  .142E+02& 37, 86&  .250E+02&  .165E+00& 37, 87&  .253E+02& -.210E+01\\
 38, 34&  .377E+00&  .182E+02& 38, 33& -.309E+00&  .164E+02& 38, 88&  .268E+02&  .225E+00& 38, 89&  .272E+02& -.201E+01\\
 39, 39&  .341E+00&  .130E+02& 39, 38& -.315E+00&  .162E+02& 39, 92&  .251E+02&  .234E-01& 39, 93&  .255E+02& -.217E+01\\
 40, 36&  .110E+00&  .181E+02& 40, 35& -.542E+00&  .163E+02& 40, 94&  .269E+02&  .828E-01& 40, 95&  .272E+02& -.209E+01\\
 41, 42&  .595E+00&  .150E+02& 41, 41& -.142E-01&  .131E+02& 41, 96&  .246E+02&  .147E+00& 41, 97&  .249E+02& -.200E+01\\
 42, 39&  .461E+00&  .149E+02& 42, 38& -.163E+00&  .181E+02& 42, 98&  .264E+02&  .202E+00& 42, 99&  .267E+02& -.193E+01\\
 43, 44&  .218E+00&  .151E+02& 43, 43& -.364E+00&  .132E+02& 43,102&  .247E+02&  .233E-01& 43,103&  .250E+02& -.207E+01\\
 44, 41&  .155E+00&  .149E+02& 44, 40& -.440E+00&  .180E+02& 44,104&  .264E+02&  .783E-01& 44,105&  .267E+02& -.199E+01\\
 45, 47&  .410E+00&  .123E+02& 45, 46& -.151E+00&  .152E+02& 45,106&  .242E+02&  .138E+00& 45,107&  .245E+02& -.191E+01\\
 46, 44&  .403E+00&  .168E+02& 46, 43& -.152E+00&  .149E+02& 46,108&  .259E+02&  .189E+00& 46,109&  .261E+02& -.184E+01\\
 47, 49&  .240E-01&  .124E+02& 47, 48& -.515E+00&  .152E+02& 47,112&  .242E+02&  .286E-01& 47,113&  .245E+02& -.197E+01\\
 48, 46&  .721E-01&  .168E+02& 48, 45& -.460E+00&  .150E+02& 48,114&  .259E+02&  .801E-01& 48,115&  .261E+02& -.190E+01\\
 49, 52&  .150E+00&  .144E+02& 49, 51& -.354E+00&  .125E+02& 49,116&  .237E+02&  .135E+00& 49,117&  .240E+02& -.183E+01\\
 50, 49&  .258E+00&  .140E+02& 50, 48& -.258E+00&  .168E+02& 50,118&  .253E+02&  .184E+00& 50,119&  .256E+02& -.176E+01\\
 51, 55&  .250E+00&  .118E+02& 51, 54& -.240E+00&  .145E+02& 51,122&  .238E+02&  .385E-01& 51,123&  .240E+02& -.187E+01\\
 52, 52&  .393E+00&  .158E+02& 52, 51& -.902E-01&  .140E+02& 52,124&  .253E+02&  .868E-01& 52,125&  .256E+02& -.181E+01\\
 53, 58&  .311E+00&  .137E+02& 53, 57& -.149E+00&  .119E+02& 53,126&  .233E+02&  .139E+00& 53,127&  .235E+02& -.174E+01\\
 54, 54&  .308E-01&  .159E+02& 54, 53& -.435E+00&  .141E+02& 54,130&  .253E+02&  .122E-03& 54,131&  .255E+02& -.185E+01\\
 55, 61&  .353E+00&  .113E+02& 55, 60& -.960E-01&  .138E+02& 55,132&  .233E+02&  .516E-01& 55,133&  .235E+02& -.179E+01\\
 56, 57&  .127E+00&  .133E+02& 56, 56& -.327E+00&  .159E+02& 56,134&  .248E+02&  .973E-01& 56,135&  .250E+02& -.173E+01\\
 57, 64&  .363E+00&  .132E+02& 57, 63& -.598E-01&  .114E+02& 57,136&  .228E+02&  .146E+00& 57,137&  .231E+02& -.166E+01\\
 58, 60&  .185E+00&  .151E+02& 58, 59& -.242E+00&  .134E+02& 58,140&  .248E+02&  .189E-01& 58,141&  .250E+02& -.177E+01\\
 59, 67&  .360E+00&  .109E+02& 59, 66& -.536E-01&  .133E+02& 59,142&  .228E+02&  .676E-01& 59,143&  .230E+02& -.170E+01\\
 60, 63&  .225E+00&  .127E+02& 60, 62& -.193E+00&  .152E+02& 60,144&  .243E+02&  .111E+00& 60,145&  .245E+02& -.165E+01\\
 61, 70&  .331E+00&  .128E+02& 61, 69& -.599E-01&  .110E+02& 61,146&  .224E+02&  .158E+00& 61,147&  .226E+02& -.159E+01\\
 62, 66&  .235E+00&  .145E+02& 62, 65& -.159E+00&  .128E+02& 62,150&  .242E+02&  .396E-01& 62,151&  .244E+02& -.168E+01\\
 63, 73&  .293E+00&  .106E+02& 63, 72& -.909E-01&  .129E+02& 63,152&  .223E+02&  .858E-01& 63,153&  .225E+02& -.163E+01\\
 64, 69&  .232E+00&  .122E+02& 64, 68& -.155E+00&  .146E+02& 64,154&  .237E+02&  .127E+00& 64,155&  .239E+02& -.157E+01\\
 65, 76&  .233E+00&  .125E+02& 65, 75& -.131E+00&  .107E+02& 65,158&  .223E+02&  .206E-01& 65,159&  .225E+02& -.166E+01\\
 66, 72&  .204E+00&  .140E+02& 66, 71& -.163E+00&  .123E+02& 66,160&  .237E+02&  .618E-01& 66,161&  .239E+02& -.161E+01\\
 67, 79&  .167E+00&  .103E+02& 67, 78& -.191E+00&  .126E+02& 67,162&  .219E+02&  .105E+00& 67,163&  .221E+02& -.155E+01\\
 68, 75&  .167E+00&  .118E+02& 68, 74& -.194E+00&  .141E+02& 68,166&  .236E+02&  .220E-02& 68,167&  .238E+02& -.163E+01\\
 69, 82&  .822E-01&  .122E+02& 69, 81& -.258E+00&  .105E+02& 69,168&  .218E+02&  .457E-01& 69,169&  .220E+02& -.158E+01\\
 70, 78&  .109E+00&  .136E+02& 70, 77& -.234E+00&  .120E+02& 70,170&  .231E+02&  .850E-01& 70,171&  .233E+02& -.153E+01\\
 71, 86&  .320E+00&  .118E+02& 71, 85& -.696E-02&  .101E+02& 71,172&  .214E+02&  .127E+00& 71,173&  .216E+02& -.148E+01\\
 72, 81&  .438E-01&  .115E+02& 72, 80& -.294E+00&  .137E+02& 72,176&  .231E+02&  .299E-01& 72,177&  .232E+02& -.156E+01\\
 73, 89&  .212E+00&  .978E+01& 73, 88& -.111E+00&  .119E+02& 73,178&  .213E+02&  .713E-01& 73,179&  .215E+02& -.151E+01\\
 74, 85&  .286E+00&  .111E+02& 74, 84& -.387E-01&  .133E+02& 74,180&  .226E+02&  .109E+00& 74,181&  .228E+02& -.146E+01\\
 75, 92&  .892E-01&  .116E+02& 75, 91& -.219E+00&  .994E+01& 75,184&  .212E+02&  .204E-01& 75,185&  .214E+02& -.153E+01\\
 76, 88&  .184E+00&  .128E+02& 76, 87& -.126E+00&  .112E+02& 76,186&  .225E+02&  .577E-01& 76,187&  .227E+02& -.149E+01\\
 77, 96&  .262E+00&  .112E+02& 77, 95& -.350E-01&  .964E+01& 77,188&  .208E+02&  .974E-01& 77,189&  .210E+02& -.144E+01\\
 78, 91&  .785E-01&  .109E+02& 78, 90& -.228E+00&  .130E+02& 78,192&  .224E+02&  .109E-01& 78,193&  .226E+02& -.151E+01\\
 79, 99&  .123E+00&  .936E+01& 79, 98& -.171E+00&  .114E+02& 79,194&  .208E+02&  .499E-01& 79,195&  .209E+02& -.146E+01\\
 80, 95&  .255E+00&  .105E+02& 80, 94& -.405E-01&  .126E+02& 80,196&  .220E+02&  .861E-01& 80,197&  .222E+02& -.142E+01\\
 81,103&  .258E+00&  .911E+01& 81,102& -.264E-01&  .111E+02& 81,200&  .207E+02&  .623E-02& 81,201&  .208E+02& -.148E+01\\
 82, 98&  .121E+00&  .122E+02& 82, 97& -.162E+00&  .107E+02& 82,202&  .219E+02&  .420E-01& 82,203&  .220E+02& -.144E+01\\
 83,106&  .963E-01&  .108E+02& 83,105& -.176E+00&  .926E+01& 83,204&  .203E+02&  .796E-01& 83,205&  .204E+02& -.139E+01\\
 84,102&  .260E+00&  .119E+02& 84,101& -.139E-01&  .104E+02& 84,208&  .218E+02&  .122E-02& 84,209&  .219E+02& -.146E+01\\
 85,110&  .199E+00&  .106E+02& 85,109& -.648E-01&  .903E+01& 85,210&  .202E+02&  .385E-01& 85,211&  .203E+02& -.141E+01\\
 86,105&  .112E+00&  .101E+02& 86,104& -.159E+00&  .120E+02& 86,212&  .214E+02&  .729E-01& 86,213&  .215E+02& -.137E+01\\
 87,113&  .276E-01&  .881E+01& 87,112& -.234E+00&  .107E+02& 87,216&  .201E+02&  .488E-03& 87,217&  .202E+02& -.143E+01\\
 88,109&  .218E+00&  .983E+01& 88,108& -.450E-01&  .117E+02& 88,218&  .213E+02&  .345E-01& 88,219&  .214E+02& -.139E+01\\
 89,117&  .103E+00&  .862E+01& 89,116& -.151E+00&  .105E+02& 89,220&  .197E+02&  .703E-01& 89,221&  .198E+02& -.135E+01\\
 90,112&  .507E-01&  .115E+02& 90,111& -.202E+00&  .997E+01& 90,222&  .209E+02&  .104E+00& 90,223&  .210E+02& -.131E+01\\
 91,121&  .163E+00&  .843E+01& 91,120& -.844E-01&  .103E+02& 91,226&  .196E+02&  .344E-01& 91,227&  .197E+02& -.136E+01\\
 92,116&  .129E+00&  .112E+02& 92,115& -.116E+00&  .973E+01& 92,228&  .207E+02&  .673E-01& 92,229&  .209E+02& -.132E+01\\
 93,125&  .208E+00&  .826E+01& 93,124& -.323E-01&  .100E+02& 93,232&  .195E+02&  .977E-03& 93,233&  .196E+02& -.138E+01\\
 94,120&  .192E+00&  .110E+02& 94,119& -.468E-01&  .951E+01& 94,234&  .206E+02&  .336E-01& 94,235&  .207E+02& -.134E+01\\
 95,128&  .574E-02&  .986E+01& 95,127& -.225E+00&  .841E+01& 95,236&  .191E+02&  .679E-01& 95,237&  .192E+02& -.130E+01\\
 96,123&  .806E-02&  .930E+01& 96,122& -.229E+00&  .111E+02& 96,240&  .205E+02&  .244E-02& 96,241&  .206E+02& -.135E+01\\
 97,132&  .314E-01&  .968E+01& 97,131& -.194E+00&  .825E+01& 97,242&  .190E+02&  .361E-01& 97,243&  .191E+02& -.131E+01\\
 98,127&  .491E-01&  .911E+01& 98,126& -.182E+00&  .109E+02& 98,244&  .201E+02&  .679E-01& 98,245&  .202E+02& -.128E+01\\
 99,136&  .452E-01&  .952E+01& 99,135& -.174E+00&  .810E+01& 99,248&  .189E+02&  .659E-02& 99,249&  .190E+02& -.133E+01\\
100,131&  .774E-01&  .893E+01&100,130& -.147E+00&  .107E+02&100,250&  .200E+02&  .378E-01&100,251&  .201E+02& -.129E+01\\
101,140&  .481E-01&  .936E+01&101,139& -.166E+00&  .795E+01&101,252&  .185E+02&  .708E-01&101,253&  .186E+02& -.125E+01\\
102,135&  .939E-01&  .876E+01&102,134& -.125E+00&  .105E+02&102,256&  .198E+02&  .100E-01&102,257&  .200E+02& -.130E+01\\
103,144&  .411E-01&  .921E+01&103,143& -.168E+00&  .782E+01&103,258&  .184E+02&  .425E-01&103,259&  .185E+02& -.126E+01\\
104,139&  .996E-01&  .860E+01&104,138& -.114E+00&  .103E+02&104,260&  .195E+02&  .728E-01&104,261&  .196E+02& -.123E+01\\
105,148&  .248E-01&  .908E+01&105,147& -.179E+00&  .770E+01&105,264&  .183E+02&  .161E-01&105,265&  .184E+02& -.128E+01\\
106,143&  .951E-01&  .845E+01&106,142& -.113E+00&  .101E+02&106,266&  .193E+02&  .461E-01&106,267&  .195E+02& -.124E+01\\
107,153&  .201E+00&  .733E+01&107,152& -.366E-03&  .894E+01&107,268&  .179E+02&  .774E-01&107,269&  .180E+02& -.120E+01\\
108,147&  .814E-01&  .831E+01&108,146& -.122E+00&  .996E+01&108,272&  .192E+02&  .210E-01&108,273&  .193E+02& -.125E+01\\
109,157&  .163E+00&  .722E+01&109,156& -.334E-01&  .882E+01&109,274&  .178E+02&  .522E-01&109,275&  .179E+02& -.122E+01\\
110,151&  .588E-01&  .818E+01&110,150& -.140E+00&  .980E+01&110,276&  .188E+02&  .813E-01&110,277&  .189E+02& -.118E+01\\
111,161&  .118E+00&  .713E+01&111,160& -.742E-01&  .870E+01&111,280&  .177E+02&  .288E-01&111,281&  .178E+02& -.123E+01\\
112,155&  .282E-01&  .806E+01&112,154& -.166E+00&  .966E+01&112,282&  .187E+02&  .576E-01&112,283&  .188E+02& -.119E+01\\
113,165&  .662E-01&  .704E+01&113,164& -.122E+00&  .859E+01&113,286&  .175E+02&  .635E-02&113,287&  .176E+02& -.123E+01\\
114,160&  .177E+00&  .927E+01&114,159& -.989E-02&  .794E+01&114,288&  .186E+02&  .352E-01&114,289&  .187E+02& -.120E+01\\
115,169&  .793E-02&  .695E+01&115,168& -.176E+00&  .849E+01&115,290&  .172E+02&  .649E-01&115,291&  .173E+02& -.117E+01\\
116,164&  .128E+00&  .915E+01&116,163& -.553E-01&  .783E+01&116,294&  .184E+02&  .142E-01&116,295&  .185E+02& -.121E+01\\
117,174&  .121E+00&  .817E+01&117,173& -.562E-01&  .687E+01&117,296&  .171E+02&  .437E-01&117,297&  .172E+02& -.118E+01\\
118,168&  .719E-01&  .903E+01&118,167& -.107E+00&  .773E+01&118,298&  .181E+02&  .715E-01&118,299&  .182E+02& -.114E+01\\

\end{tabular} 
\end{table}
\nopagebreak

      As can be seen from stellar evolution, nuclear fusion powers stars. The conversion of hydrogen into helium involves a chain of reactions, called the PPI chain, which have that conversion as their net product. The reactions involved are $H^1+H^1=D^2+\beta^++\nu$, $D^2+H^1=He^3+\gamma$ and $He^3+He^3=He^4+2H^1$. H and He have always been present with the ratio of H:He of roughly 9:1 because of the PPI chain. The rest of the elements, including C, N, O, Mg, Si and Fe are all made in the stars by the nuclear fusion process. In this process, most of the lighter new elements have been made. However, fusion can only produce elements up to mass number of 56 and atomic number of 26. Heavier nuclides being less stable than nuclides with mass number of 56, cannot be spontaneously produced by fusion. Therefore, the heavier nuclides are produced by neutron capture and proton capture processes during the regular evolution of stars or during the supernova explosion. One aspect of changing magic numbers is its influence on the rapid neutron capture process or the r-process path of nucleosynthesis where the neutron capture proceeds on a rapid time scale as compared to the $\beta$-decay lifetimes. The r-process follows a path at the extreme neutron rich side of the valley of stability with neutron separation energies close to 1.2-2.0 $MeV$ (close to the neutron drip line). One of the characteristics of the path is that the r-process goes through neutron magic numbers. Thus, if new magic numbers appear in the neutron-rich region, the r-process path may be modified. Such modification might solve inconsistencies between calculations and observations for the r-process abundances. For the unstable nuclei, when the time between successive neutron captures is much larger than the $\beta$-decay lifetimes, the network of processes involved is called the slow neutron capture process or the s-process and closely follows the valley of $\beta$-stability. Because proton capture requires overcoming an energy barrier, it is not an efficient process. Therefore, the p-process nuclides have low abundances compared to those of the s and r nuclei. But some proton rich nuclei can not be synthesized by either the s-process chains or the decay of neutron rich matter and are synthesized by the astrophysical proton capture (p) process nucleosynthesis responsible for few proton rich stable nuclei and via the rapid-proton capture (rp) process which follows a path close to the proton drip line highlighting its astrophysical importance.  

\section{Summary and conclusion}
\label{section4}

      Neutron and proton separation energies have been calculated using the newly modified Bethe-Weizs\"acker mass formula which delineates the positions of all new and old magic numbers. The last bound isotopes of neutron rich nuclei have been identified using this new BW mass formula and the new neutron and proton drip lines have been provided. The drip line nuclei listed and the values of the separation energies provided in the table could be useful for planning experiments on mass measurements. The even-odd staggering has been found to be quite prominent for the proton dripline showing that the proton dripline for even Z nuclei is further away from the region of $\beta$-stability than that for odd Z nuclei. However, the even-odd staggering has not been found to be strong for the neutron dripline. Since the new mass formula identifies all the new magicities or its loss, calculations according to this mass formula provide better limits to the neutron and proton drip lines.

\end{document}